\documentclass[aip,jcp,reprint,amsmath,amssymb]{revtex4-2}
\usepackage{graphicx}
\usepackage{xcolor}
\usepackage{amsmath}
\usepackage{soul}

\begin{document}

\title{Programming patchy particles to form three-dimensional dodecagonal 
quasicrystals}

\author{Daniel F.\ Tracey}
\affiliation{Physical and Theoretical Chemistry Laboratory, Department of Chemistry, University of Oxford, Oxford, OX1 3QZ, United Kingdom}
\author{Eva G. Noya}
\affiliation{Instituto de Qu\'{i}mica F\'{i}sica Rocasolano, Consejo Superior de Investigaciones Cient\'{i}ficas, CSIC, Calle Serrano 119, 28006 Madrid, Spain}
\author{Jonathan P.\ K.\  Doye}
\email{jonathan.doye@chem.ox.ac.uk}
\affiliation{Physical and Theoretical Chemistry Laboratory, Department of Chemistry, University of Oxford, Oxford, OX1 3QZ, United Kingdom}

\date{\today}

\begin{abstract}
Model patchy particles have been shown to be able to form a wide variety of structures, including symmetric clusters, complex crystals and even two-dimensional quasicrystals. Here, we investigate whether we can design patchy particles that form three-dimensional quasicrystals, in particular targeting a quasicrystal with dodecagonal symmetry that is made up of stacks of two-dimensional quasicrystalline layers. We obtain two designs that are able to form such a dodecagonal quasicrystal in annealing simulations. The first is a one-component system of 7-patch particles but with wide patches that allow them to adopt both 7- and 8-coordinated environments. The second is a ternary system that contains a mixture of 7- and 8-patch particles, and is likely to be more realizable in experiments, for example, using DNA origami. One interesting feature of the first system is that the resulting quasicrystals very often contain a screw dislocation.
\end{abstract}

\maketitle 

\section{\label{sec:intro}Introduction}

Quasicrystals are characterized by long-range order (exemplified by sharp Bragg peaks in their diffraction patterns) in the absence of translational periodicity, and often exhibit symmetries not feasible in periodic crystals. The initial examples were found in metallic alloy systems with the most common symmetries being icosahedral\cite{Shechtman84} and decagonal,\cite{Bendersky85} but with metastable octagonal\cite{Wang87} and dodecagonal\cite{Ishimasa85} quasicrystals also being observed. More recently, an increasing number of soft-matter quasicrystals have been discovered,\cite{Dotera11} nearly all of which are dodecagonal; these include polymeric micelles,\cite{Zeng04,Fischer11,Zhang12d} and nanoparticle superlattices.\cite{Talapin09,Ye17}

Why do quasicrystals, rather than periodic crystals, sometimes form, and what are the key features of the interparticle interactions that cause this? Theory and simulations can play an important role in answering these questions. Quasicrystals have been most commonly observed in simulations for isotropic potentials with multiple features in the potential whose positions are tuned to favour quasicrystalline order.\cite{Dzugutov93,Engel07,Dotera14,Pattabhiraman15,Engel15,Ryltsev15,Damasceno17,Ryltsev17} For ultrasoft potentials (i.e.\ particles without a hard repulsive core), it can be shown more rigorously that quasicrystals with a particular symmetry can be favoured when the ratio of two inverse length scales in the Fourier transform of the potential has a specific value.\cite{Subramanian16,Savitz18,Ratliff19} 

Simulation examples that are exceptions to this ``two-length-scale'' stabilization mechanism include quasicrystals formed from hard particles (tetrahedra\cite{HajiAkbari09,Je21} and triangular bipyramids\cite{HajiAkbari11b} can form dodecagonal quasicrystals) and from ``patchy'' particles with directional bonding;\cite{vanderLinden12,Reinhardt16,Noya21} the latter are the focus of this paper. In particular, two-dimensional patchy particles with five and seven regularly arranged patches have been found to form dodecagonal quasicrystals that are square-triangle tilings where the bonds are equally likely to be oriented along twelve equivalent directions.\cite{vanderLinden12} 

The 5-patch system has become a model system to better understand quasicrystal behaviour. For example, phase diagrams have been computed, showing that the dodecagonal quasicrystal is the thermodynamically stable phase for an intermediate range of temperature and patch width, and its stabilization is due to its greater entropy compared to crystalline approximants.\cite{Reinhardt13b} The growth dynamics of this quasicrystal have also been studied.\cite{Gemeinhardt18} 

\begin{figure}
\includegraphics[width=3.3in]{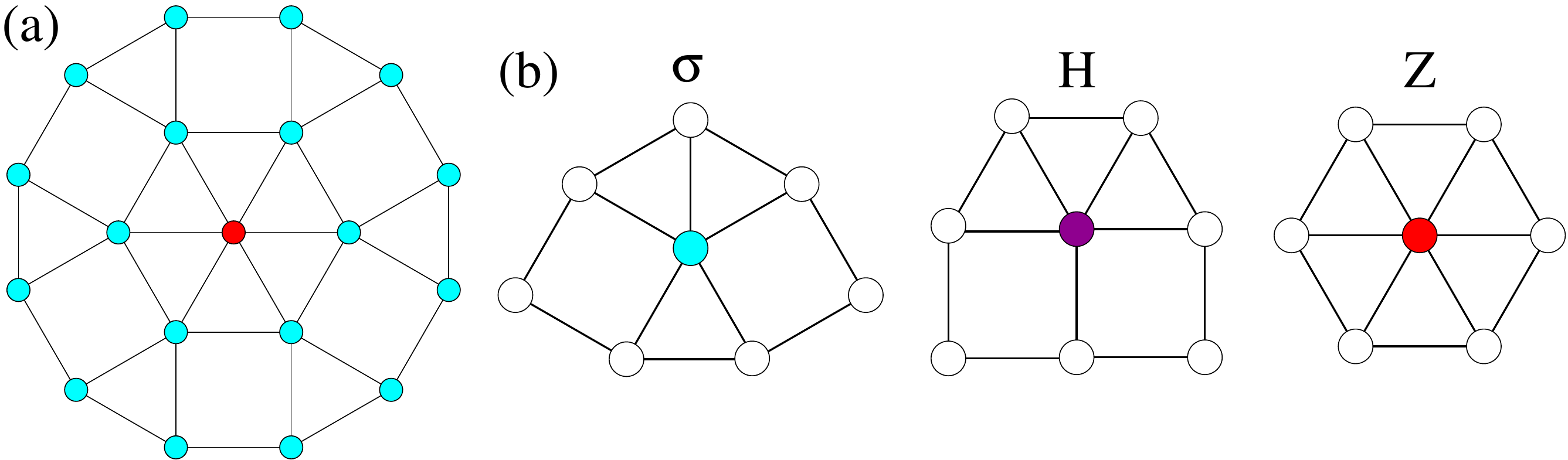}
\caption{\label{fig:2D} (a) A dodecagonal motif that is a key motif in two-dimensional dodecagonal quasicrystalline square-triangle tilings. (b) Three possible local motifs in square-triangle tilings.
The labels mirror those used for Frank-Kasper phases.\cite{FrankK58,FrankK59}}
\end{figure}

A key structural motif in these two-dimensional dodecagonal quasicrystals is the dodecagon shown in Fig.\ \ref{fig:2D}(a). The 5- and 6-coordinate environments in the dodecagon are both feasible when the particles' patches are sufficiently wide. To realize such a quasicrystal, for particles with instead a fixed maximum number of interaction partners, requires a mixture of 5- and 6-patch particles,\cite{Reinhardt16} this being achieved experimentally in systems of multi-arm DNA tiles on a surface.\cite{Liu19}

Here, we extend this approach to obtain 3D dodecagonal quasicrystals, exploring the additional complexities that the added dimension brings. It is noteworthy that many of the dodecagonal quasicrystals previously observed in simulations have been for 2D\cite{Engel07,vanderLinden12,Dotera14,Pattabhiraman15,Reinhardt16} or quasi-2D\cite{Metere16,Damasceno17} systems.

\section{\label{sec:methods}Methods}

\subsection{\label{sec:potential}Potential}

The interaction between particles is described using the patchy-particle model introduced in Refs.\ \onlinecite{Wilber09b,Tracey19}. In this pair potential, the interaction is described by a Lennard-Jones repulsive core and an attractive tail modulated by angular and torsional dependent functions:
\begin{widetext}
\begin{equation}
    V_{ij}(\mathbf{r}_{ij},\mathbf{\Omega}_i,\mathbf{\Omega}_j) = \begin{cases}
                V_{\mathrm{LJ}}^\prime(r_{ij}) & : r_{ij} < \sigma_{\mathrm{LJ}}^{\prime} \\
                V_{\mathrm{LJ}}^\prime(r_{ij}) \underset{{\mathrm{patch~pairs~}\alpha,\beta}}{\max} \left[ \varepsilon_{\alpha\beta}V_{\mathrm{ang}}(\mathbf{\hat{r}}_{ij},\mathbf{\Omega}_i,\mathbf{\Omega}_j)V_{\mathrm{tor}}(\mathbf{\hat{r}}_{ij},\mathbf{\Omega}_i,\mathbf{\Omega}_j) \right] & : r_{ij} \geqslant \sigma_{\mathrm{LJ}}^{\prime}
                \end{cases},
                \label{eq:V}
                \end{equation}
\end{widetext}
where $\mathbf{r}_{ij}$ is the interparticle vector,
$\alpha$ and $\beta$ are patches on particles $i$ and $j$, respectively, $\mathbf{\Omega}_i$ is the orientation of particle $i$, $V_{\mathrm{LJ}}^\prime(r)$ is a cut-and-shifted Lennard-Jones (LJ) potential and
$\sigma_{\mathrm{LJ}}^{\prime}$ 
corresponds to the distance at which $V_{\mathrm{LJ}}^\prime$ passes through zero. We set the cutoff distance $r_\mathrm{cut}=2.5\,\sigma_\mathrm{LJ}$. $\varepsilon_{\alpha\beta}$ is a measure of the relative strength of the interactions between patches $\alpha$ and $\beta$; $\varepsilon_{\alpha\beta}$ is set to zero for patches that do not interact.

The {angular} modulation term $V_\mathrm{ang}$ is a measure of how directly the patches $\alpha$ and $\beta$ point at each other, and is given by
        \begin{equation}
        V_\mathrm{ang}(\mathbf{\hat{r}}_{ij},\mathbf{\Omega}_i,\mathbf{\Omega}_j) = \exp \left( -\frac{\theta_{\alpha ij}^{2}}{2\sigma_{\mathrm{ang}}^{2}} \right) \exp \left( -\frac{\theta_{\beta ji}^{2}}{2\sigma_{\mathrm{ang}}^{2}} \right) ,
        \label{eq:Vang}
        \end{equation}
$\theta_{\alpha ij}$ is the angle between the patch vector $\mathbf{\hat{P}}_{i}^{\alpha}$, representing the patch $\alpha$, and $\mathbf{\hat{r}}_{ij}$. $\sigma_\mathrm{ang}$ is a measure of
the angular width of the patch.

The {torsional} modulation term $V_\mathrm{tor}$ describes the variation in the potential as either of the particles is rotated about the interparticle vector $\mathbf{r}_{ij}$ and is given by
\begin{equation}
        V_{\mathrm{tor}}(\mathbf{\hat{r}}_{ij},\mathbf{\Omega}_i,\mathbf{\Omega}_j)
        = \exp\left(
                - \frac{1}{2\sigma_{\mathrm{tor}}^{2}}\left[
                                \min\limits_{\phi^{\mathrm{offset}}_{\alpha\beta}}
                                \left(\phi_{\alpha\beta}-\phi^{\mathrm{offset}}_{\alpha\beta} \right)
                        \right]^{2}
        \right).
        \label{eq:Vtor}
\end{equation}
where $\phi^\mathrm{offset}$ is the preferred value of the torsional angle $\phi$.
To define the torsional angle $\phi_{\alpha\beta}$, a unique {reference vector} (here, this is always one of the other patch vectors) is associated with each patch.
In order to capture the symmetry of an environment, more than one equivalent offset angle can be defined; in these cases we find the minimum value of $\phi-\phi_\mathrm{offset}$ across the set of equivalent offset angles. 
To turn off the torsional component of a particular interaction we set $V_\mathrm{tor}=1$.
We use $\sigma_\mathrm{tor}= 2\sigma_\mathrm{ang}$ throughout.

We use $\sigma_\mathrm{LJ}$ as our unit of length, and the LJ well depth $\varepsilon_\mathrm{LJ}$ as our unit of energy. Values of $\sigma_\mathrm{ang}$ and $\sigma_\mathrm{tor}$ are given in radians. 
Temperatures are given in reduced form, $T^*=k_B T/\varepsilon_\mathrm{LJ}$. Full details of each patchy particle model that we study are tabulated in the Supporting Information.

\subsection{\label{sec:simulations}Simulations}

We use Metropolis Monte Carlo to simulate the patchy-particle systems in the canonical ensemble. The diffusive dynamics this algorithm generates is appropriate for modelling the dynamics of colloidal particles in solution. To simulate the larger systems we use a GPU-enabled code that uses the parallelization strategy introduced in Ref.\ \onlinecite{Anderson13}. Like in Ref.\ \onlinecite{Tracey19}, assembly is achieved by slow annealing simulations in which the temperature is linearly decreased with time. The system is initiated from a disordered, low-density ($\rho=0.1 \sigma^{-3}_{\mathrm{LJ}}$) fluid state at a temperature just above that for which nucleation of a condensed-state cluster can occur on the simulation time scales.

\subsection{\label{sec:characterize}Structural characterization}

We analyse the structure of the clusters that grow in a number of ways. To identify the different local particle environments we use a common-neighbour analysis,\cite{Clarke93,Reinhardt16} focussing on those environments that are relevant to the target dodecagonal quasicrystals. In particular, we identify the three-dimensional equivalents of the local environments in Fig.\ \ref{fig:2D}(b), where the in-plane bonding is the same as in the figure, but with out-of-plane bonding to particles directly above and below the particle of interest. Environments not matching these three are labelled as U (unidentified); this category will include surface particles with incomplete bonding and particles at defect sites. 

To identify quasicrystallinity, we compute diffraction patterns for our assembled structures. Dodecagonal quasicrystallinity is indicated by a 12-fold symmetric pattern when viewed from a direction perpendicular to the quasicrystalline layers. 
Specifically, we compute the real part of the structure factor $S(\mathbf{q})$:
\begin{equation}
S(\mathbf{q})=\frac{1}{N} \sum_{j=1}^{N}\sum_{k=1}^{N}e^{-i\mathbf{q}\cdotp\mathbf{r}_{kj}}.
\end{equation}
We define the normal to the layers as the average direction of the axial patches on the particles in a cluster, and, after orienting this direction along $z$, we evaluate the structure factor for $\mathbf{q}=(q_x,q_y,0)$. We evaluate the diffraction patterns averaged over a set of configurations at the end of  our annealing runs. 

\section{\label{sec:results}Results}

The basic design approach that we use 
is to target structures that are stacked crystalline or quasicrystalline layers, where atoms in adjacent layers lie directly above each other. The patch designs that mediate interactions within the layers will be similar to the previous two-dimensional examples.\cite{vanderLinden12,Reinhardt16} In addition, axial patches perpendicular to the above mediate interactions between the layers. The torsional component of the interactions facilitates the layer formation. Specifically, for the reference vector of the equatorial patches we use one of the axial patches with preferred offset angles of 0 or 180$^\circ$.

\begin{figure}
\includegraphics[width=3.3in]{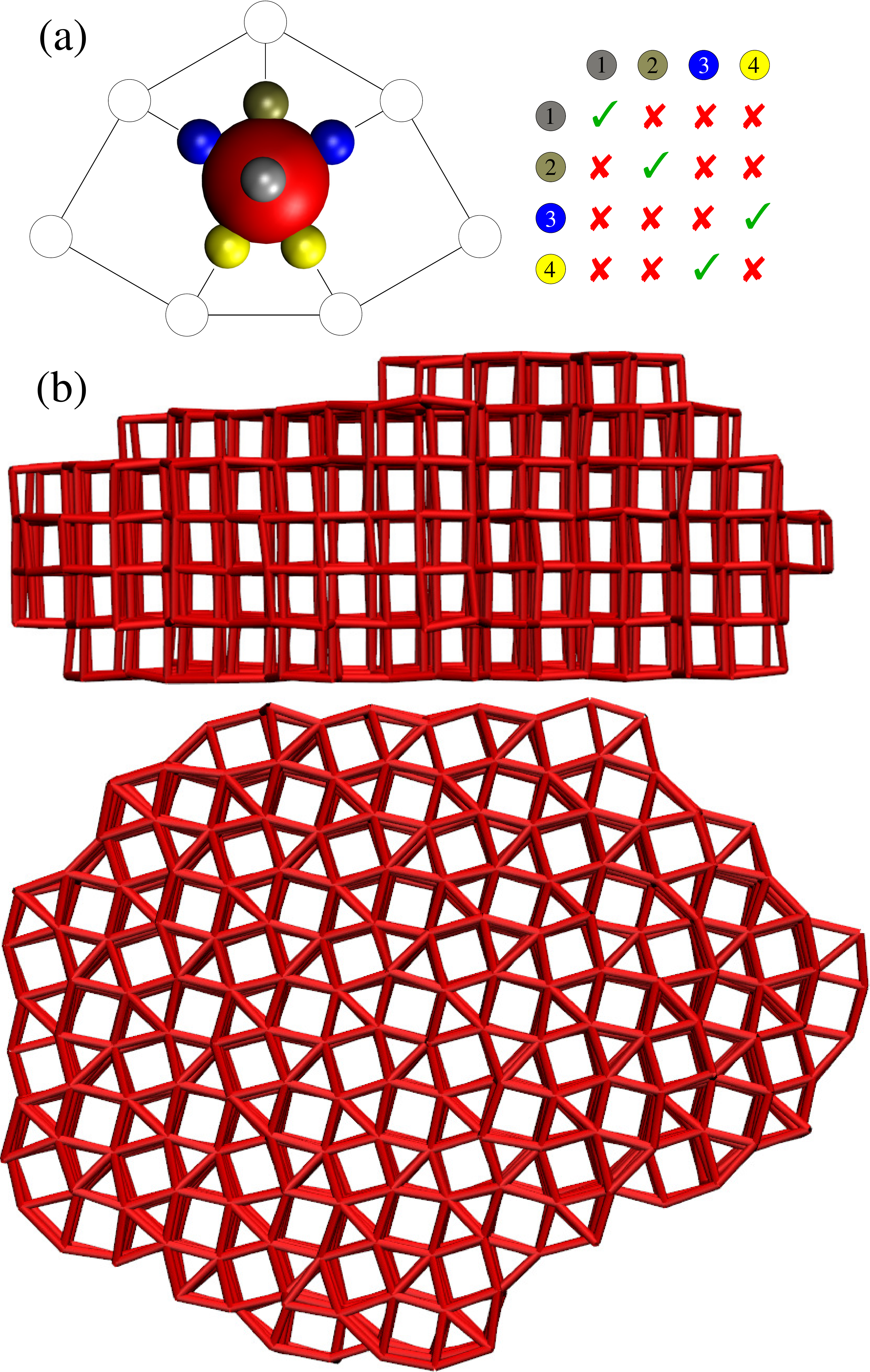}
\caption{\label{fig:sigma}Stacked $\sigma$-crystal. (a) The patchy particle design and its relationship to the $\sigma$ local environment, and the interaction matrix between the different types of patches. 
(Note there is a second axial patch of type 1 (silver) on the opposite side of the particle shown). 
(b) Two views of an 864-particle crystalline cluster grown in an annealing simulation. Only the bond network is visualized. The stacked layers and the in-plane $\sigma$-like order are both apparent.
$\sigma_\mathrm{ang}=0.3$. 
}
\end{figure}

\subsection{\label{sec:sigma}Stacked-$\sigma$ crystal design}

We first illustrate our approach for a crystal made of stacked ``$\sigma$-layers'', where each particle has an in-plane environment identical to the $\sigma$ environment in Fig.\ \ref{fig:2D}(b). The crystal is tetragonal with space group $P4/mbm$ (127) with the $4g$ Wyckoff sites occupied. As each environment in the crystal is equivalent, only one particle type is needed. Three types of equatorial patches are used for the three unique types of in-plane bonds, with the inter-patch angles exactly matching those of the $\sigma$ environment, i.e.\ they are 60$^\circ$ or 90$^\circ$ if they are part of in-plane triangles or squares, respectively (Fig.\ \ref{fig:sigma}(a)).
This design fully specifies the target crystal \cite{Tracey19} and, as expected, the crystal forms straightforwardly in annealing simulations for moderately narrow patches (we use $\sigma_\mathrm{ang}=0.3$) where 7-fold coordination 
is clearly favoured. 
An example crystallite is illustrated in Fig.\ \ref{fig:sigma}(b); note the anisotropic crystal shape reflects
the greater stabilization provided by the larger number of in-plane bonds.

\subsection{\label{sec:unary}One-component quasicrystal design}

\begin{figure*}
\includegraphics[width=6.7in]{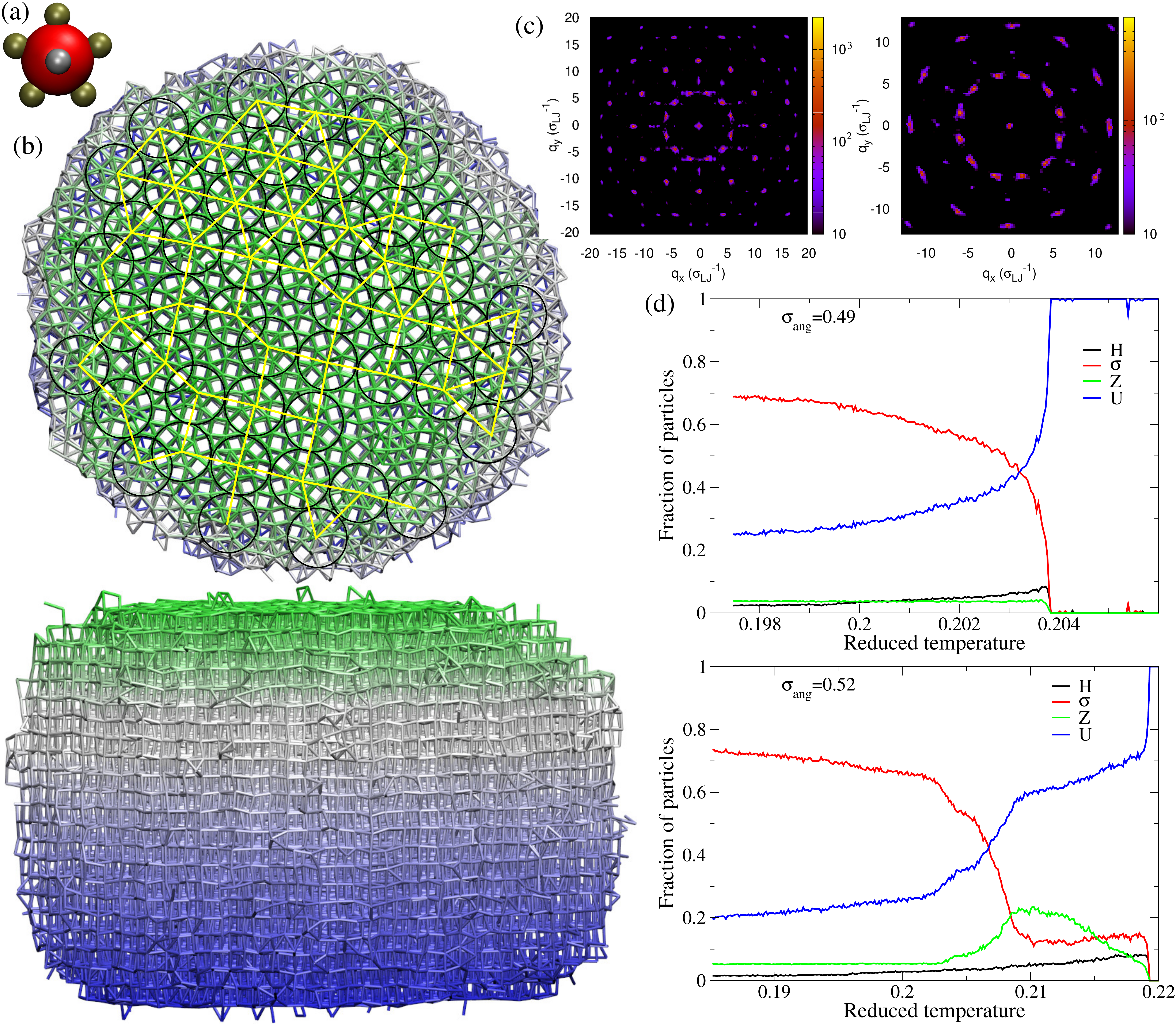}
\caption{\label{fig:unary} One-component dodecagonal quasicrystal assembly. 
(a) The patchy particle design. 
(b) An example final configuration from an annealing run at $\sigma_\mathrm{ang}=0.52$, viewed from the side and above. The graduated coloring is to add visual clarity. 
In the top view each dodecagon is highlighted by a black circle, and yellow lines connect the centres of edge-sharing or overlapping dodecagons.
(c) Example diffraction patterns for the above configuration (left) and for a cluster from another annealing runs at the same $\sigma_\mathrm{ang}$ (right) viewed perpendicular to the stacked layers. Each diffraction pattern is averaged over a set of configurations at the end of the annealing run.
(d) The fraction of particles in the largest cluster with a given local environment as a function of annealing temperature in a system of 20\,000 particles at $\sigma_\mathrm{ang}=0.49$ and 0.52. The annealing rate for the temperature was 1.2 and $3.87 \times 10^{-4}$ per $10^5$ MC cycles, respectively. Further properties for these two annealing runs are shown in Supplementary Fig.\ S2.
}
\end{figure*}

We next consider a potential one-component quasicrystal-forming design. The particle design is very similar to the above, but all five equatorial patches are now equivalent and regularly-spaced in the equatorial plane, i.e.\ the in-plane inter-patch angles are 72$^{\circ}$ (Fig.\ \ref{fig:unary}(a)).

We first consider assembly at $\sigma_\mathrm{ang}=0.3$ where we expect a strong preference for 7-fold coordination (5-fold coordination in the layers); $\sigma$ environments are more likely than H environments because the $\sigma$ environment has a slightly better match to the inter-patch angles.\cite{vanderLinden12}
The crystallites resulting from our annealing simulations are very similar to those in Section \ref{sec:sigma}, but with some particles in H environments (see Supplementary Fig.\ S1).

The more interesting behaviour is anticipated for larger $\sigma_\mathrm{ang}$, because, as the potential becomes less anisotropic, an increasing propensity for 6-fold in-plane coordination is expected. For the previously studied 2D systems, this leads to a small region of stability for the dodecagonal quasicrystal in the ($\sigma_\mathrm{ang},T$) phase diagram, sandwiched between regions where the fluid, $\sigma$ crystal or hexagonal crystal is most stable.\cite{Reinhardt13b} On annealing, two types of assembly pathways were seen for this 2D system depending on the value of $\sigma_\mathrm{ang}$. At lower $\sigma_\mathrm{ang}$, the quasicrystal assembled direct from the fluid, whereas at slightly larger $\sigma_\mathrm{ang}$, the system first formed a hexagonal crystal before transforming into a quasicrystal at lower temperature.

We find somewhat similar behaviour in the annealing simulations for our 3D
systems, observing both direct and indirect pathways for quasicrystal assembly.
For example, at $\sigma_\mathrm{ang}=0.49$ a quasicrystalline cluster grows
directly from the fluid with the order increasing as the temperature decreases
(Fig.\ \ref{fig:unary}(d)). By contrast,  at $\sigma_\mathrm{ang}=0.52$ the
cluster that first assembles from the gas phase has a much higher proportion of
hexagonal in-plane environments and unidentified environments and a much lower
fraction of $\sigma$ environments. The cluster in this temperature regime
exhibits significant disorder both within the layers and in terms of a reduced
tendency to form planar layers. However, at low temperature the cluster then
transforms into a quasicrystal with the final cluster having a very similar
distribution of environments to the quasicrystals that grow direct from the gas
phase. 

Diffraction patterns for the final clusters from two of the annealing runs are shown in Fig.\ \ref{fig:unary}(c); they have clear 12-fold symmetry confirming the quasicrystalline character. From the top view of this configuration (Fig.\ \ref{fig:unary}(b))
the planar layers can be seen to be tilings of squares and triangles, with the different local environments and the dodecagonal motifs of Fig.\ \ref{fig:2D} evident. Although there is no translational order, orientational order is maintained through the whole system with the inter-plane bonds equally likely in twelve equivalent directions. The dodecagonal motifs are often seen to share edges, forming larger motifs such as square or triangular arrangements of dodecagons; also present in the structure are examples where 
the dodecagons overlap. The network of interconnected dodecagons propagates through the whole structure.

\begin{figure*}
\includegraphics[width=6.7in]{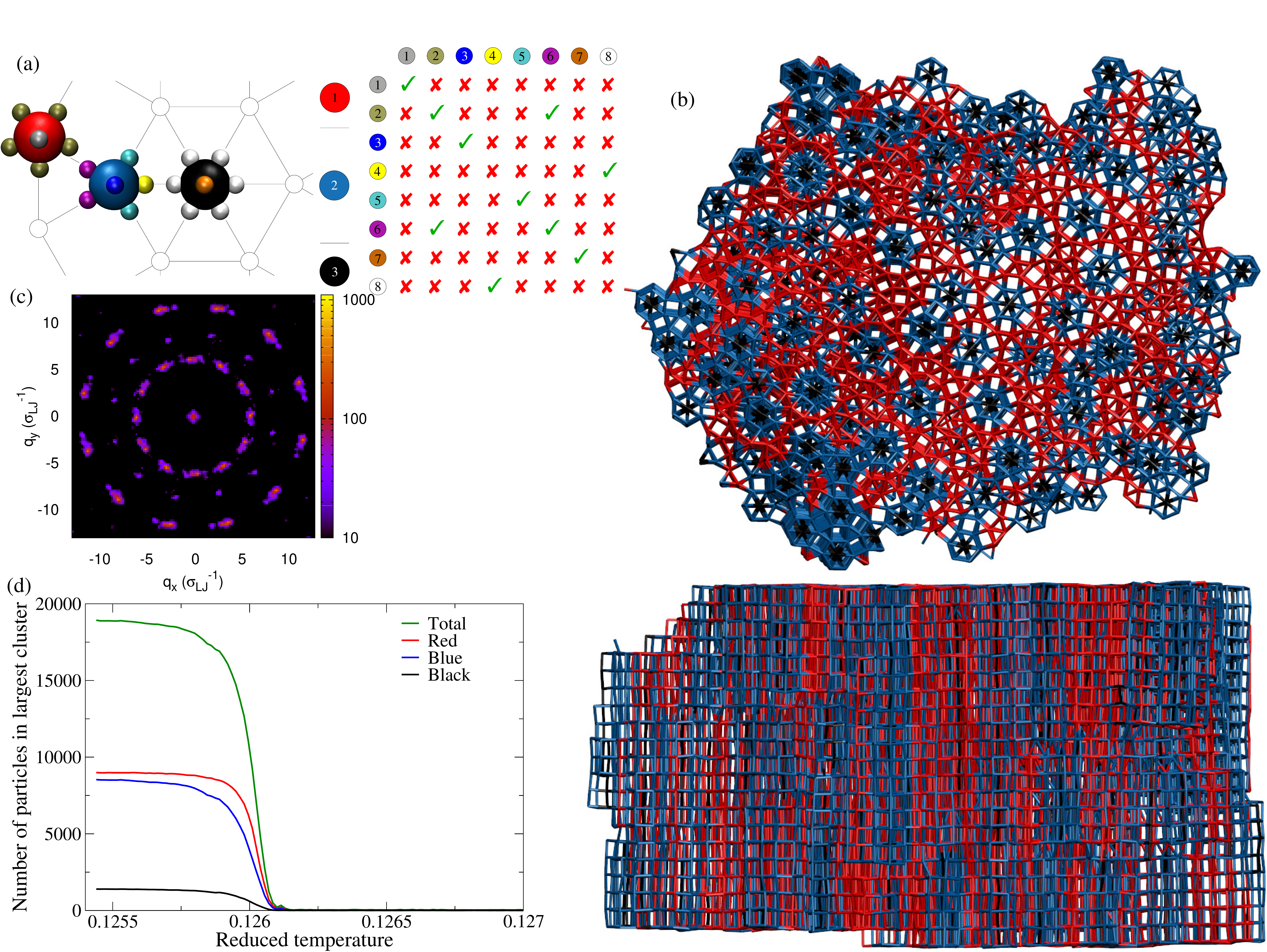}
\caption{\label{fig:ternary} Ternary dodecagonal quasicrystal assembly. 
(a) The three types of patchy particles, their relationship to the basic dodecagonal motif, and the interaction matrix.
(b) An example final configuration from an annealing run at $\sigma_\mathrm{ang}=0.3$ for a system of $N=20,007$ particles. The relative proportions of particle types 1, 2 and 3 were 6/13, 6/13 and 1/13, respectively. The annealing rate for the temperature was $5\times 10^{-5}$ per $10^5$ MC cycles. The colors match the particle types in (a).  
(c) The diffraction pattern for this configuration viewed perpendicular to the stacked layers.
(d) The number of particles in the largest cluster broken down by particle type. Snapshots of the growth of this quasicrystalline cluster are given in Supplementary Figure S3.
}
\end{figure*}

\subsection{\label{sec:ternary}Ternary quasicrystal design}

One disadvantage of the above one-component system is that the quasicrystal only forms for a relatively small range of $\sigma_\mathrm{ang}$ in which multiple coordination environments compete as the potential becomes less anisotropic. Instead, it is probably experimentally easier to realize patchy particles with well-defined valence. Therefore, we also explored quasicrystal formation for multi-component designs where the environments with different coordination numbers are achieved by particles with different numbers of patches. These systems are somewhat analogous to the 2D examples studied in Ref.\ \onlinecite{Reinhardt16}. Such systems have the advantage that quasicrystal formation is likely to be relatively insensitive to the precise value of the patch width; we choose to use $\sigma_\mathrm{ang}=0.3$; this is in the regime where directional bonding is dominant. However, multi-component designs also have potential disadvantages. Having multiple components is likely to have some adverse affects on the kinetics. Assembly is likely to be somewhat slower as particles must search for their correct partners, and rearrangements of the particles when in the condensed state may be significantly slower due to the greater selectivity in the bonding. Also, there is the potential that the particles might prefer to phase-separate rather than form an ordered phase that involves all particle types.

After exploratory testing of different models, the most promising design that we then studied in more detail is given in Fig.\ \ref{fig:ternary}(a). The system has three particle types. Particle type 3 has six in-plane patches and is designed to be at the centre of the dodecagonal motif. The other two particle types have five in-plane patches. One (type 2) is designed to bind around the eight-patch particle to form the hexagonal motif at the centre of the dodecagon. The other (type 1) is geometrically equivalent to the particle studied in the above one-component example, with all equatorial patches equivalent, and will typically be located on the outside of the dodecagonal motifs. Two patch types can bind to multiple partners. This flexibility allows the particles to organize in multiple ways, reducing the tendency for crystalline structures to form. The magenta patches on particle type 2, as well as bonding to the equatorial patches of particle 1 can also self-interact. The latter allows overlapping dodecagonal motifs to form. Similarly, the equatorial patches on particle type 1, as well as binding to the magenta patches on particle type 2, can also self-interact.

Thus, particle type 1 can potentially form a fully-bonded $\sigma$-crystal on its own, and particle types 2 and 3 could together form a crystal of overlapping dodecagons without particle type 1. Therefore, this system has the potential to demix rather than form an ordered ternary phase. In the previous sections, all attractive patch-patch interactions were of equal strength (i.e.\ $\varepsilon_{\alpha\beta}=1$). 
Here, we used the values of $\varepsilon_{\alpha\beta}$ to tune the behaviour of the system to prevent phase separation. In particular, we reduced the strength of the self-interactions of patches 2 and 6 by 10\% (i.e.\ $\varepsilon_{22}=\varepsilon_{66}=0.9$ and $\varepsilon_{26}=1$), to slightly favour bonding between particle types 1 and 2. Without this change, the initial clusters tended to predominantly involve particles 1, with particles 2 and 3 only being added to the outside of the clusters once the concentration of particles 1 in the fluid phase had significantly diminished.

In this model we also reduced the strength of the axial interactions on all particles by 5\% (i.e.\ $\varepsilon_{11}=\varepsilon_{33}=\varepsilon_{77}=0.95$). This design choice was simply to change the shape of the assembled clusters, so that they had a greater lateral extent, thus allowing easier identification of the potential quasicrystalline order. As in Ref.\ \onlinecite{Reinhardt16}, the composition was chosen to match that for a triangular crystal of edge-sharing dodecagonal units.

Fig.\ \ref{fig:ternary}(b) shows one of the clusters resulting from our annealing simulations, and its  diffraction pattern. Again, the diffraction pattern has clear 12-fold order confirming its quasicrystalline character (Fig.\ \ref{fig:ternary}(c)). Examination of the top view of the configuration shows that the planes are square-triangle tilings, with dodecagons being a common motif although somewhat less prevalent than for the one-component system. While the system clearly naturally forms a ternary cluster --- the hexagonal motifs are distributed throughout the cluster --- the composition seems to be slightly higher in particle type 1 towards the centre of the planes. This tendency is also evident in Fig.\ \ref{fig:ternary}(b); the number of particles of type 2 in the growing cluster somewhat lags that of type 1, even though there are equal numbers in the simulation. The initial nucleation event always occurred around a small hexagonal stack (Supplementary Fig.\ S3).

\begin{figure*}
\includegraphics[width=6.7in]{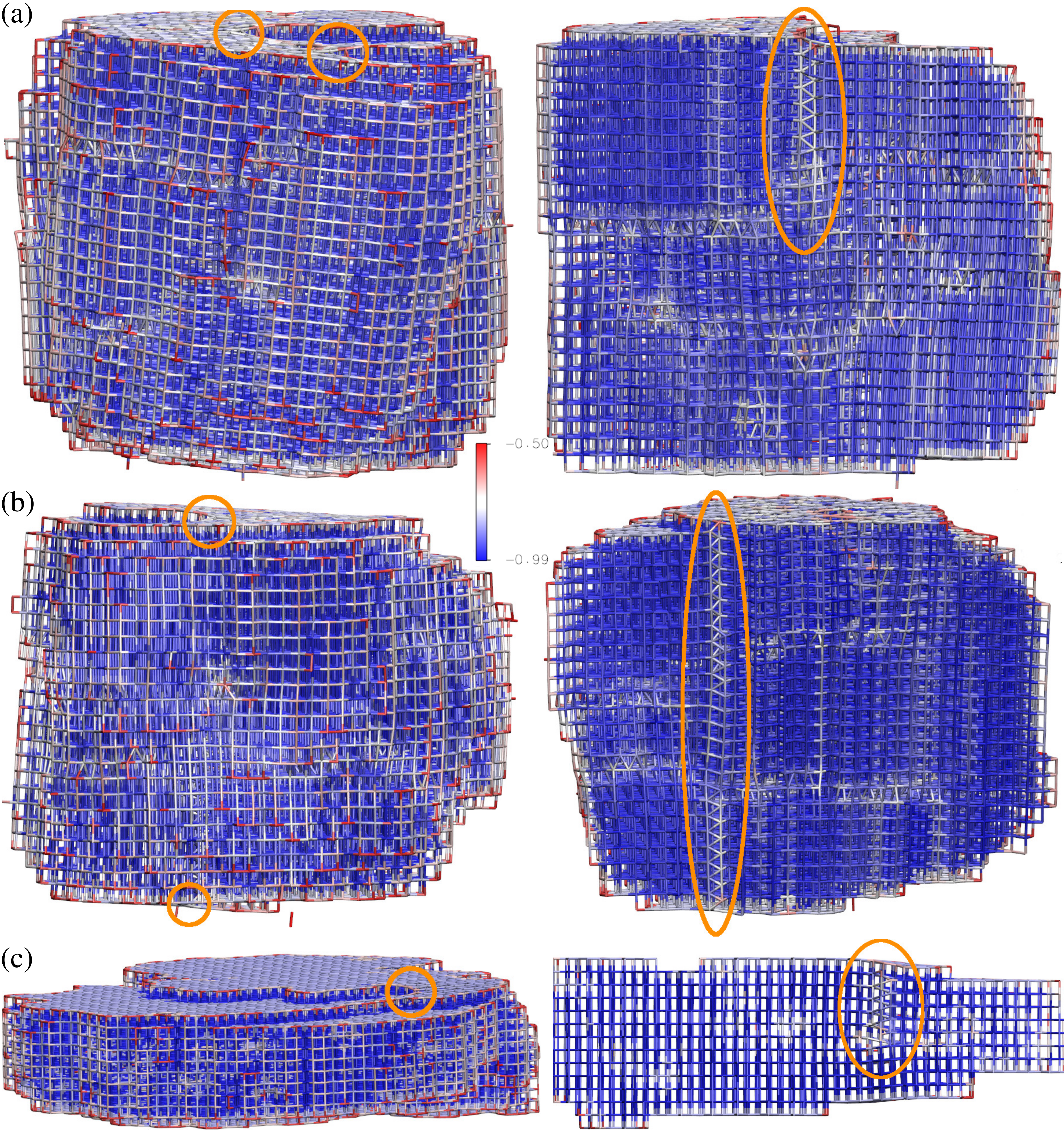}
\caption{\label{fig:screw} Configurations exhibiting screw dislocations for one-component quasicrystals formed at (a) $\sigma_\mathrm{ang}$=0.43, and (b) $\sigma_\mathrm{ang}$=0.46, and (c) for the stacked $\sigma$-crystal at $\sigma_\mathrm{ang}$=0.3. On the left are the complete clusters with the exit points
of the screw dislocations highlighted by orange circles. On the right are sections through the clusters that allow the path of the dislocation through the structures to be seen (highlighted by the orange ellipses). These configurations were prepared by taking the final configuration from an annealing simulation, and then quenching it to $T^*=0.01$ to remove thermal noise. The particles are colored according to their potential energy per patch.}
\end{figure*}

\subsection{\label{sec:screw}Screw dislocations}

An especially interesting feature of the assembled one-component quasicrystals was the ubiquitous presence of screw dislocations. They were present in the final configurations for all clusters assembled at
$\sigma_\mathrm{ang}=0.43$ and 0.46, and many
at 0.49 and 0.52. 
Their lower likelihood of occurrence at larger $\sigma_\mathrm{ang}$ suggests their formation is somewhat suppressed when the quasicrystal does not form direct from the fluid, but rather via an intermediate condensed state. 

Two examples are shown in Fig.\ \ref{fig:screw}. The screw dislocation runs through the clusters roughly perpendicular to the quasicrystalline planes and leads to a ledge on the top surface of the quasicrystal at the exit point of the dislocation. During growth the ledge spirals around the defect core, leading to continuous growth without the need for the secondary nucleation of an island on top of a flat layer. For this reason, screw dislocations are associated with enhanced crystal growth rates.

In a crystal, the Burgers vector is parallel to the screw dislocation and perpendicular to the plane of the Burgers circuit. By analogy, the Burgers vectors in our quasicrystals would be in the axial direction perpendicular to the quasicrystalline layers. However, defining a Burgers circuit in a quasicrystal is less straightforward, and the formal theory of dislocations in quasicrystals makes use of higher-dimensional crystallography.\cite{Sandbrink14}

To aid the visualization of the dislocations, the final configurations from the annealing simulations were quenched to reduce thermal noise (by simulating the systems at $T^*=0.01$ for $5\times 10^5$ MC cycles). The particles were then colored according to their potential energy. This leads to expected differences in color between particles in bulk, surface or edge sites, but also allows the path of the screw dislocation to be easily identified in cuts through the structures. This color scheme also highlights other defects that are present in the bulk of the structures, but these are not our focus here. 

The first example in Fig.\ 5(a) has two screw dislocations emerging on the top surface. Consequently, there is a noticeable tilt to the layers. The cut clearly shows the helical pattern of bonds around the dislocation core. The stress associated with the dislocations is quite localized, with clear deviations from the bulk energy only noticeable close to the dislocation core. The screw dislocations are generally very straight but with occasional jogs or kinks. Thus, the screw dislocation in the cut in Fig.\ 5(a) is only in the plane of the cut for the top third of the cluster, whereas for the example in Fig.\ 5(b) it is visible passing through the entire structure. 

Why do screw dislocations occur for the one-component quasicrystal but not the ternary example? It is probably for two reasons. Firstly, the wide patches that we use in the one-component case (to facilitate multiple environments) give rise to smaller energetic penalties for the deformed bonds around the dislocation core. When we performed simulations for this system with $\sigma_\mathrm{ang}=0.3$, we never observed screw dislocations. By contrast, when we simulated larger systems of the stacked $\sigma$ crystal at $\sigma_\mathrm{ang}=0.3$, screw dislocations sometimes formed (Fig.\ \ref{fig:screw}(c)). In this case, the more anisotropic shape of the crystals may be relevant as the total energetic cost of the dislocation will be proportional to the crystal thickness, thus partially offsetting the greater local cost of the screw dislocation. Secondly, it may be harder to create a pattern of bonding at a dislocation core that is compatible with the greater specificity of the interactions in the ternary case.

Another important question is why and at what stage do the screw dislocations form. Clearly, for systems of this size, it is very unlikely that screw dislocation formation is thermodynamically favourable, because the entropy gain from their presence is unlikely to outweigh their unfavourable energy. For the same reason, they are unlikely to be present in pre-critical nuclei which are in quasi-equilibrium with the fluid. This is especially so given the critical nucleus size is likely to be small at the temperatures at which we see spontaneous nucleation in our simulations. Therefore, it is likely that they become incorporated into growing post-critical clusters. The question then is what happens when, during the growth, a local structural perturbation forms on the surface of the cluster that could potentially lead to a fully-formed screw dislocation: does it get annealed out because of its less favourable energy, or does it get incorporated and locked in due to the boost it provides to the growth kinetics? Under some conditions the latter seems to happen sufficiently frequently that the resulting quasicrystals invariably include screw dislocations.

\begin{figure}
\includegraphics[width=3.3in]{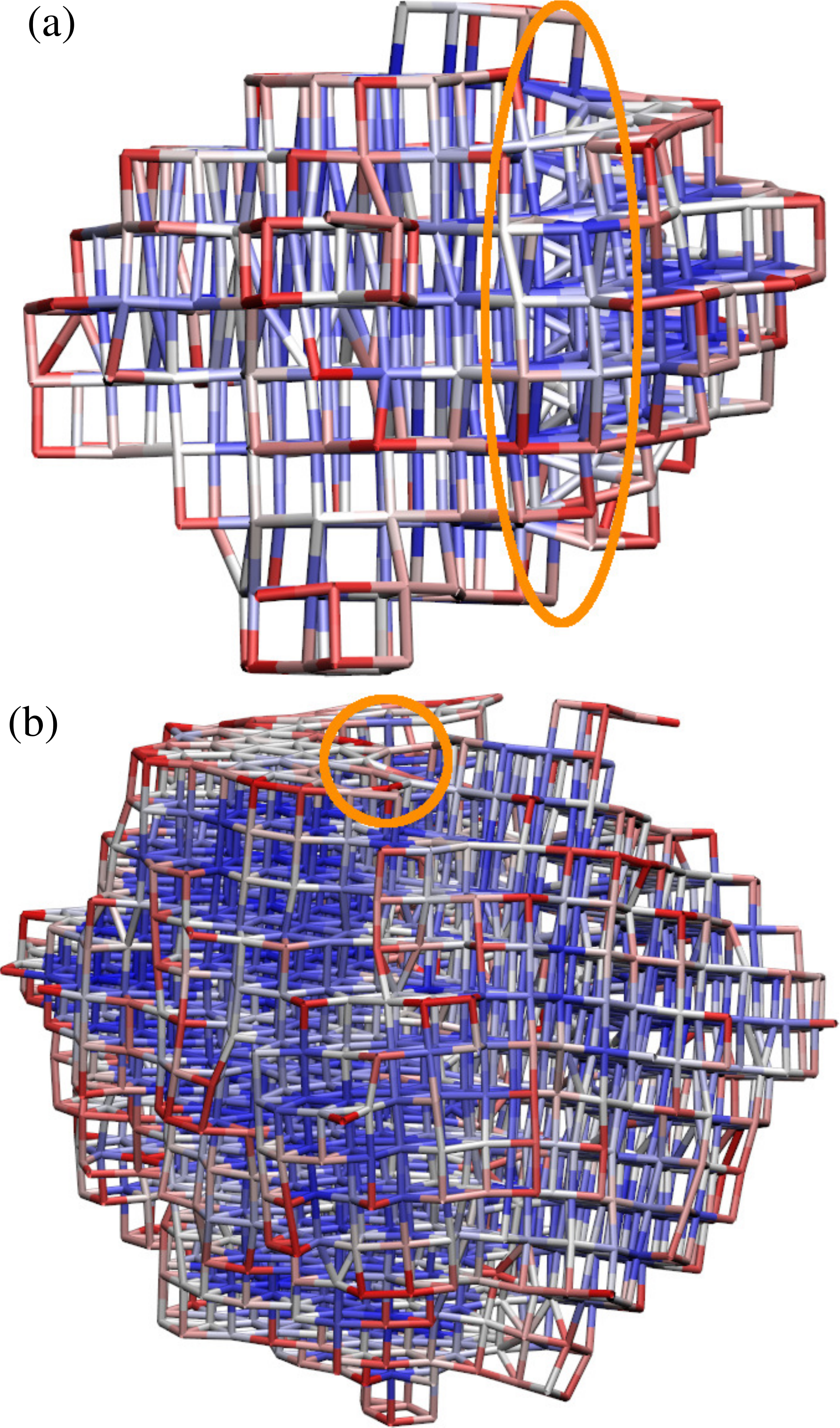}
\caption{\label{fig:screw_evolve} Configurations from the early stages of cluster growth that show the presence of screw dislocations. (a) $N\approx 540$ at $\sigma_\mathrm{ang}$=0.46. (b) $N\approx 1990$ at $\sigma_\mathrm{ang}$=0.43. As in Fig.\ \ref{fig:screw}, the cluster configuration was quenched to $T^*=0.01$ to remove thermal noise, and the particles colored according to their potential energy per patch.
}
\end{figure}

Analysis of the growing clusters indicates that the screw dislocations generally form in the early stages of cluster growth. Fig.\ \ref{fig:screw_evolve}(a) shows a cluster with just over 500 particles that has a screw dislocation  near to the edge of the cluster. The larger cluster in Fig.\ \ref{fig:screw_evolve}(b) (around 2\,000 particles) already has two screw dislocations passing through the complete cluster. It is likely that as the clusters grow larger it becomes less likely that new screw dislocations are incorporated, 
because of the increased formation costs due to the greater thickness of the clusters.

Even once fully formed, the screw dislocations can potentially be removed if they diffuse to the edge of the clusters and are then expelled. However, dislocation motion was slow on our computational time scales, so we rarely saw this occur.

\section{\label{sec:conclusion}Conclusions}

We have shown that patchy particles can be programmed to form 3D dodecagonal quasicrystals. We have introduced two designs that achieve this. The first is a one-component system with wide patches that enable the particles to adapt to multiple environments, in particular ones with different coordination numbers. The second is a ternary mixture, where particles with seven and eight patches explicitly match the different coordination environments in the quasicrystal; the well-defined valence of these particles is likely to make them easier to realize experimentally. 
In this system, the patch-patch interaction strengths had to be tuned in order to prevent phase separation of the ternary mixture. The quasicrystallinity of both systems was confirmed by the 12-fold symmetry of their diffraction patterns. 

One interesting feature of the one-component systems was that the majority of the quasicrystals that we grew had screw dislocations passing through the structures in the axial direction. The features of this system that facilitated their formation were the wide patches and the layer-like structure. The screw dislocations were incorporated into the quasicrystals in the early stages of growth. In these instances, when the initial structural perturbation at the cluster surface that could potentially lead to a screw dislocation occurred, the tendency for the screw dislocation to facilitate growth must have won out over the tendency for the defect to be annealed out because of its less favourable potential energy. It is perhaps noteworthy that, although there have been many simulations of screw dislocations,\cite{Maresca18} we are unaware of any simulation studies where screw dislocations have spontaneously appeared during crystal growth. 

The quasicrystals that have been observed for soft matter systems have almost invariably been of dodecagonal symmetry.\cite{Dotera11} Thus, it would be particularly interesting if patchy particle designs could be developed that form three-dimensional quasicrystals with other symmetries. We have now very recently achieved this for icosahedral quasicrystals, where the target quasicrystalline structures were generated by projection from six-dimensional cubic lattices.\cite{Noya21} Generating appropriate target structures for other axial quasicrystals, however, may be challenging. In this study, the task was made easier firstly because we chose the simplest way one could conceive of making a 3D axial quasicrystal, namely by a simple stacking of two-dimensional quasicrystalline layers. Secondly, the square-triangle character of the dodecagonal quasicrystalline layers is straightforwardly compatible with a particle model because the edges of the tiling are all of the same length. Thirdly, we also already knew that two-dimensional patchy particles could form such quasicrystalline layers.\cite{vanderLinden12,Reinhardt16}

Experimental examples for which the structure of the quasicrystal has been solved may provide one source of target structures.\cite{Steurer09} However, the high coordination numbers typically present in metallic alloys may present additional challenges for a patchy-particle approach. Another potential source of target structures is from simulation studies where 3D axial quasicrystals have formed in particle-based models with complex isotropic potentials. These include dodecagonal quasicrystals that have a Frank-Kasper-like particle decoration of square and triangular prismatic units,\cite{Dzugutov93,Ryltsev17} two decagonal examples\cite{Ryltsev15,Damasceno17} and one octagonal example.\cite{Damasceno17}

How might such patchy particles be realized experimentally? Although there has been much progress in synthesizing colloidal analogues of patchy particles,\cite{Glotzer07,Li20,Liu20} the degree of control over the geometry of the patches is not yet sufficient to realize particles like those envisaged here. An alternative might be to use the techniques of DNA nanotechnology.\cite{Ma20} For example, DNA origami polyhedra\cite{Veneziano16} whose vertices have been decorated with short DNA single strands that facilitate inter-vertex binding have been used to form a variety of simple crystal structures.\cite{Tian20} For this system, the inter-vertex binding is quite flexible, but we have also proposed an approach in which a six-helix bundle extends from each vertex to provide more directional and torsionally specific interactions between the DNA origami.\cite{Noya21} The sequences of the single strands that allow inter-origami binding can be tuned to provide the requisite specificity, and it would be relatively straightforward to choose the type of polyhedron and tune their edge lengths to match the geometry of the patchy particle. As the origami particles would have a well-defined maximum number of binding partners, they would be more appropriate for realizing the ternary quasicrystal. For example, in this scheme, a hexagonal bipyramidal DNA origami would be used to create an analogue of particle type 3 with its six in-plane and two out-of-plane patches, and pentagonal bipyramidal origamis would be used for particle types 1 and 2.

\begin{acknowledgments}
D.F.T. is grateful for funding via the ESPRC Centre for Doctoral Training in
Theory and Modelling in Chemical Sciences, under grant EP/L015722/1.
E.G.N. acknowledges funding from the Agencia Estatal de
Investigaci\'{o}n (AEI) and the Fondo Europeo de Desarrollo Regional
(FEDER) under grant numbers FIS2015-72946-EXP(AEI) and
FIS2017-89361-C3-2-P(AEI/FEDER,UE), and the European Union’s Horizon
2020 research and innovation programme under the Marie
Sk\l{}odowska-Curie grant agreement No.\ 734276. 
We acknowledge the use of the University of Oxford Advanced Research Computing (ARC) facility and the Cambridge Service for Data Driven Discovery (CSD3).
\end{acknowledgments}

\section*{Data Availability}
The data that support the findings of this study are available from the corresponding author upon reasonable request.


\end{document}